\documentclass[a4paper,11pt]{article}
\pdfoutput=1
\usepackage{jheppub}
\usepackage[T1]{fontenc}
\usepackage{cancel,tensor,enumitem,fix-cm}
\usepackage{amsthm}
\usepackage{amssymb}

\newcommand{\be}{\begin{equation}}
\newcommand{\ee}{\end{equation}}
\newcommand{\bea}{\begin{eqnarray}}
\newcommand{\eea}{\end{eqnarray}}

\renewcommand{\Re}{\operatorname{Re}}

\subheader{\begin{flushright}
\end{flushright}}

\title{{\fontsize{7.5mm}{9mm}\selectfont Decoherence and thermalization of Unruh-DeWitt detector in arbitrary dimensions}}

\author{Hao Xu}
\affiliation{Center for Gravitation and Cosmology, College of Physical Science and Technology, Yangzhou University, 180 Siwangting Road, Yangzhou City, Jiangsu Province  225002, China}
\emailAdd{haoxu@yzu.edu.cn}

\abstract{We study the decoherence and thermalization of an Unruh-DeWitt detector linearly coupled to the free massless scalar field in flat spacetime with arbitrary dimensions $d\geq 2$. The initial state of the detector is chosen to be a pure state consisting of a linear superposition of ground and excited states, and we calculate the time evolution of reduced density matrix of the detector. Using perturbation method, we analytically derive the transition rate of the detector (the rate of change of the diagonal elements in the density matrix) and the decoherence rate (the rate of change of the off-diagonal elements in the density matrix). We find that the results are not the same in odd and even dimensional spacetimes, but the unitarity of the qubit is preserved in both cases. The real part of the decoherence rate is related to the transition rate, while the imaginary part may contain different forms of divergence terms in different dimensions due to the temporal order product operator and the singularities of the Wightman function for quantum field theory. We derive the recurrence formula to obtain the divergence terms in each dimension and analyze the renormalization problem.
}

\begin{document}

\maketitle

\flushbottom

\section{Introduction}

The relationship between general relativity and quantum physics has been one of the most significant open topics in modern physics. Despite the belief that there should exist a more fundamental theory of quantum gravity which degenerates into general relativity and quantum physics in certain situations, the pursuit of quantum gravity has not been very successful. It appears that there are considerable differences between the fundamental laws regulating macroscopic and microscopic processes. For example, Einstein's equation, the fundamental equation of general relativity, is nonlinear, whereas Schrödinger's equation, the fundamental equation of quantum mechanics, is linear. Because of the linearity of Schrödinger's equation, the linear superposition of the two solutions is still a solution of the equation, leading to the principle of superposition and a probabilistic interpretation of our universe in the macroscopic world. We cannot help but wonder: if the universe is governed by fundamental quantum mechanics principles at the microscopic level, why does the macroworld appear to be classical?

One of the reasons for this problem is that we used the incorrect assumption of idealized closed system. Although the Schrödinger equation is linear, the process by which we measure quantum systems inevitably introduces extra interactions \cite{RevModPhys.76.1267}. In fact, a purely closed system is not even physically possible, since we would not be able to study it and thus can not obtain any information about it. No physical system, including the observer, can be completely isolated from its surroundings, thus the theory of open quantum systems is the relevant theoretical framework for our research \cite{Breuer}, and any study or measurement is based on the investigation of some subsystems of the open system. When a quantum system interacts with its environment as a subsystem, the time evolution of the total system can be unitary, but if we consider only the subsystem, its evolution is not unitary. The interaction with the environment leads to the loss of quantum coherence (a measure of the definite phase relation between different states of the system). We refer to it as decoherence, which can be thought of as the loss of quantum nature caused by the fact that each system is loosely coupled to its environment \cite{Zeh1970,Zurek1981,Zurek1982,Zurek2003}. Decoherence is also an important factor in explaining how the classical world emerges from the quantum regime, where the concept of ``classicality'' is understood as an emergent concept in the dynamical description of the quantum-to-classical transition.

On the other hand, gravity is the most fundamental and universal interaction among macroscopic phenomena, and there have even been suggestions that gravity may be the ultimate explanation for the collapse of the wave function \cite{Diosi:1988uy,Penrose:1996cv}. It is of great physical importance to study gravitationally induced decoherence. We refer to all decoherence due to gravitational effects as gravitational decoherence. See for example \cite{Bassi:2017szd,Anastopoulos:2021jdz} for reviews and \cite{Petruzziello2022} for recent developments in this area. Of course, the effect of gravity is a very broad description. Spacetime fluctuates, and the fluctuations may be classical, quantum, or both. Classical fluctuations are perturbations of spacetime geometry, such as the gravitational wave. Quantum fluctuations take into account the quantization of gravity itself, which is still an open problem, and predictions still depend on assumptions about the behavior of gravity at the quantum scale. In fact, the quantized spacetime itself decoheres, so that the classical spacetime structure emerges from quantum gravity, which is also sometimes called gravitational decoherence \cite{Zurek1986,Kiefer1992}. In addition, gravity can also affect other physical systems, causing these systems to exhibit new features, such as quantum field theory in curved spacetime \cite{Davies}. These new features are also likely to affect decoherence.

Here we give a brief discussion of quantum field theory in curved and flat spacetime, since it will be the focus of the present work. In quantum field theory, particles are interpreted as excitations of the Fock basis, and energy eigenstates are collections of particles with definite momenta. In flat spacetime, the notion of a particle is characterised mathematically by time translation Killing vector, and the field equation will have plane-wave solutions of definite frequency that extend throughout space. However, a generic curved spacetime need not admit any timelike Killing vectors and we may not even find such plane-wave solutions, so the definition of a particle is not intuitive. The solution to this problem is to think operationally and define a sensible notion of a particle detector that reduces to our intuitive picture in flat spacetime, then the particle can be defined as ``what the particle detectors detect''\cite{Davies1984}. Particle detectors give us a more intuitive way of studying gravity and quantum field theory, as the observer is directly present and the system can be studied in an open system approach.

The Unruh-DeWitt detector is the most basic type of particle detector \cite{DeWitt1979}. It is an idealised point-like object with internal energy levels. It was originally proposed to study the Unruh effect, which states that for the uniformly accelerated observer the vacuum of quantum fields in Minkowski spacetime is transformed into the thermal state \cite{Fulling:1972md,Unruh:1976db,Unruh:1983ms}. Although it occurs in flat spacetime, the Unruh effect teaches us the most important lesson of quantum field theory in curved spacetime: the idea that ``vacuum'' and ``particles'' are observer-dependent rather than fundamental concepts. Since the causal structure of Rindler space describing the motion of an accelerated observer is similar to the maximally extended Schwarzschild spacetime describing an eternal black hole, we can also obtain Hawking radiation directly from the Unruh effect. Although there are many features that we can only explore in more detail using a curved metric, the Unruh effect still captures the essence of quantum field theory in curved spacetime, and thus deserves further study.

In the present work, we consider the decoherence and thermalization of the Unruh-DeWitt detector in different dimensions due to the Unruh effect. The Unruh-DeWitt detector is assumed to be a qubit with two energy levels: the ground state $|g\rangle $ and the excited state $|e\rangle $. When the qubit is accelerated in flat spacetime, because of its interaction with the vacuum state of quantum field theory, the qubit with initial state $|g\rangle$ can then be excited to the state $|e\rangle$, and the population satisfies the Boltzmann distribution corresponding to a temperature proportional to the acceleration. Previous studies have focused more on the population and heat exchange between the detector and the quantum field theory in different backgrounds, e.g.\cite{Fukuma:2013uxa,Rabochaya:2015aza,Ng:2016hzn,Hotta:2020pmq,Arias:2017kos,Gray:2018ifq,Xu:2019hea,Ng:2018drz,Ahmed:2020fai,Ahmed:2023uem,Pitelli:2021oil,Hodgkinson:2014iua} and references therein, while in this work we are more interested in the decoherence of the detector.

We assume that the initial state of the qubit is a pure state, consisting of a linear superposition of $|g\rangle $ and $|e \rangle $, thus the off-diagonal elements of the initial density matrix of the qubit are nonzero. The qubit is coupled linearly to the free massless scalar field in flat spacetime of arbitrary dimensions ($d\geq 2$), and we calculate the time evolution of the reduced density matrix of the qubit. Using the perturbation method, we can derive the transition rate (the rate of change of the diagonal elements in the reduced density matrix) and the decoherence rate (the rate of change of the off-diagonal elements in the reduced density matrix) analytically to second order. The transition rate and the real part of the decoherence rate are easier to compute, and they \emph{do not} evolve independently but are related. Although the results are not the same in odd and even dimensional spacetimes, the unitarity of the qubit is always preserved. On the other hand, the imaginary part of the decoherence rate may contain different forms of divergence terms in different dimensions due to the temporal order product operator and the singularities of the Wightman function for quantum field theory. We will derive the recurrence formulas of different dimensions, obtain the form of the divergence terms in each dimension, and analyze the renormalization problem.

The remainder of the paper is organized as follows. In section \ref{section2} we present our model and calculate the transition rate and decoherence rate by perturbation method. Using the periodicity of Wightman function, we can directly obtain the transition rate in any dimension. We will also present Theorem 1, which gives the real part of the decoherence rate and leads to a Corollary showing that the unitarity is always preserved in the time evolution of Unruh-DeWitt detector in all dimensions. In section \ref{section3} we compute in detail the cases $d = 2, 3$ and prove that the imaginary part of the decoherence rate is finite in $d = 2, 3$. In section \ref{section4} we give the recurrence formula for general $d$ and Theorem 2, which predicts the imaginary part of the decoherence rate in any dimensions, especially the divergent terms. In section \ref{section5} we give a brief summary of our main results and close with conclusions.

\section{Transition rate and decoherence rate in general $d$\label{section2}}

The total Hamiltonian of the qubit(Unruh-DeWitt detector)-field system can be written as
\begin{equation}
\hat{H}_{\text{total}}=\hat{H}_0+\hat{H}_{\text{int}},
\end{equation}
and the $\hat{H}_0$ is the free Hamiltonian of qubit and quantum field theory
\begin{equation}
\hat{H}_0=\hat{H}+\hat{H}_{\text{field}},
\end{equation}
where  $\hat{H}=\frac{\omega}{2}{\hat{\sigma}_z}$ ($\hat{\sigma}_z$ is the Pauli matrix and $\omega$ is the energy gap between $|g\rangle $ and $|e \rangle $) and $\hat{H}_{\text{field}}$ is the Hamiltonian associated to the free massless scalar field in $d$-dimensional spacetime. The qubit-field interaction Hamiltonian is given by
\begin{equation}
\hat{H}_{\text{field}} = \lambda \hat{\sigma}_x \phi[x(\tau)],
\end{equation}
where $\lambda$ is a small coupling constant of the interaction, $\hat{\sigma}_x=|g\rangle \langle e|+|e\rangle \langle g|$ is the qubit monopole operator which allows population to be exchanged between energy levels, and $\phi[x(\tau)]$ is the scalar field evaluated at the spacetime point where the qubit is located. The initial state for the quantum field theory is vacuum state $|0\rangle$.  For the qubit, we choose it to be $\sqrt{1-p}|g\rangle+\sqrt{p}|e\rangle$, in which both $\sqrt{1-p}$ and $\sqrt{p}$ are real numbers, thus the initial density matrix of the qubit is
\begin{equation}
\rho_{\text{0}}=\begin{pmatrix} 
    p & \sqrt{p(1-p)} \\

    \sqrt{p(1-p)} & 1-p
\end{pmatrix}.
\end{equation}
The inital state of the total system is then $\rho_{\text{in}}=\rho_{\text{0}}\otimes |0\rangle \langle 0|$.

In order to solve the time evolution of the system, we move to the interaction picture. The quantum Liouville’s equation can be written as
\begin{equation}
i\frac{\mathrm{d} \rho^{\text{I}}(t)}{\mathrm{d}t}=[\hat{H}_{\text{int}}^{\text{I}},\rho^{\text{I}}(t)],
\end{equation}
where we used the ``I'' to denote the operators in the interaction picture. The solution is given in terms of a Dyson series:
\begin{equation}
\rho^{\text{I}}(t)=\rho_{\text{in}}-i\int_0 ^t \mathrm{d}\tau[\hat{H}_{\text{int}}^{\text{I}}(\tau),\rho_{\text{in}}]-\frac{1}{2}\int_0 ^t \mathrm{d}\tau \int_0 ^t \mathrm{d}\tau' T\left\{\left[\hat{H}_{\text{int}}^{\text{I}}(\tau),[\hat{H}_{\text{int}}^{\text{I}}(\tau'),\rho_{\text{in}}]\right]\right\}+\cdots,
\end{equation}
where $T$ is the temporal order product operator. The $\rho^{\text{I}}(t)$ is the density matrix of the total system. In order to obtain the reduced density matrix of the qubit, we also have to take partial trace over the basis of quantum field theory. Since the field operator $\phi[x(\tau)]$ is contained in $\hat{H}_{\text{int}}^{\text{I}}(\tau)$, the contribution of the $n$-th order perturbation term comes from the $n$-point correlation function of the quantum field theory. For the free massless scalar field, if the $n$ is odd the correlation function vanishes, and since the coupling constant $\lambda$ is small, we will consider up to second order in perturbation theory. After some tedious calculation, we can have the perturbation of the reduced density matrix of qubit is

\begin{eqnarray}
\delta \rho^{\text{I}}(t) &=&\lambda^2 \int_0 ^t \mathrm{d}\tau \int_0 ^t \mathrm{d}\tau'G^+_d(\Delta \tau) \Bigg[\begin{pmatrix} 
    (1-p)e^{-i\omega \Delta \tau}-pe^{i\omega \Delta \tau} & \sqrt{p(1-p)}e^{i\omega(\tau+\tau')} \\

    \sqrt{p(1-p)}e^{-i\omega(\tau+\tau')} & pe^{i\omega \Delta \tau}-(1-p)e^{-i\omega \Delta \tau}
\end{pmatrix}\nonumber \\
&-&\theta(\Delta \tau)\sqrt{p(1-p)}\begin{pmatrix} 
    0 & e^{i\omega \Delta \tau} \\

    e^{-i\omega\Delta \tau} & 0
\end{pmatrix}-\theta(-\Delta \tau)\sqrt{p(1-p)}\begin{pmatrix} 
    0 & e^{-i\omega \Delta \tau} \\ 

    e^{i\omega\Delta \tau} & 0
\end{pmatrix}
\Bigg], 
\end{eqnarray}
where $G^+_d(\Delta \tau)=\langle 0 |\phi[x(\tau)]\phi[x(\tau')]|0\rangle$ is the positive frequency Wightman function of the free massless scalar field in $d$-dimensional spacetime and $\Delta \tau\equiv \tau-\tau'$. Defining a new variable $T=\frac{\tau+\tau'}{2}$ \footnote{This $T$ should not be confused with temporal order product operator.}, we can have the transition rate as 
\begin{equation}\label{tr}
\frac{\partial \rho^{\text{I}}_{11}}{\partial T}=-\frac{\partial \rho^{\text{I}}_{22}}{\partial T}=\lambda^2 \int_{-\infty} ^{+\infty} \mathrm{d} (\Delta \tau) G^+_d(\Delta \tau)\left( (1-p)e^{-i\omega \Delta \tau}-pe^{i\omega \Delta \tau}\right),
\end{equation}
and the decoherence rate
\begin{equation}\label{dr}
\frac{\partial \rho^{\text{I}}_{12}}{\partial T}=\frac{\partial \rho^{\text{I}*}_{21}}{\partial T}=-\lambda^2  \sqrt{p(1-p)} \left(\int_{0} ^{+\infty }\mathrm{d} (\Delta \tau)G^+_d(\Delta \tau)e^{i\omega \Delta \tau}+\int_{-\infty} ^{0}\mathrm{d} (\Delta \tau)G^+_d(\Delta \tau)e^{-i\omega \Delta \tau}\right).
\end{equation}
These two formulas is the first main result of this work. We can easily find that the integration range of transition rate is from $-\infty$ to $+\infty$, while decoherence rate covers only half due to the temporal order product operator. 

Next we consider the positive frequency Wightman function of the free massless scalar field in $d$-dimensional spacetime. For $d\geq 3$, the $G^+_d(\Delta \tau)$ takes the form of 
\begin{equation}
G^+_d(\Delta \tau)=\frac{\Gamma(d/2-1)}{4\pi^{d/2}}z^{2-d},
\label{Gd}
\end{equation}
where $z=\varepsilon+i\Delta \text{sgn}(t-t')$, $\varepsilon$ is an infinitesimal positive quantity of dimension of length, and $\Delta=[-(x-x')^2]^{1/2}$ denotes the seperation. In the case of $d=2$, the above formula is divergent, and we can use dimensional regularization to obtain
\begin{equation}
G^+_2(\Delta \tau)=\lim_{d\rightarrow 2}G^+_d(\Delta \tau)=-\frac{1}{2\pi}\log z+\lim_{d\rightarrow 2}\frac{\Gamma(d/2-1)}{4\pi^{d/2}}.
\end{equation}
The infinite constant is due to the infrared divergence inherent in the two-dimensional massless field. For the Unruh-DeWitt detector, we have the world line
\begin{equation}
t=\frac{1}{a}\sinh(a\tau),\qquad  x=\frac{1}{a}\cosh(a\tau),
\end{equation}
so we can get
\begin{eqnarray}
z=\frac{2i}{a}\sinh \frac{a}{2}(\Delta \tau-i\varepsilon).
\end{eqnarray}
Now we have the positive frequency Wightman function for the Unruh-DeWitt detector in the free massless scalar field for all $d\geq2 $-dimensional spacetime.

With the general form of $G^+_d(\Delta \tau)$, in principle, for each $d$ we can put it into the eq.\eqref{tr} and \eqref{dr}, and thus obtain the transition rate and the decoherence rate. The key to the problem here is the integration of the complex function. In many cases, we can use the Cauchy-Riemann residue theorem to evaluate the integral by expressing it as a limit of the contour integral. For any $d$, the $G^+_d(\Delta \tau)$ has infinite singularities (including the one at $i\varepsilon$) with periodicity ${2\pi i}/{a}$ on the vertical axis of the complex plane, and we need to add up all the contributions from each pole.

In this work we will try to give a general result and proof. In complex analysis, the integral from $-\infty$ to $+\infty$ is much easier to calculate, so we consider the transition rate eq.\eqref{tr} first. Here we follow the derivation in \cite{Takagi:1986kn}. If we define
\begin{equation}
F_d(\omega)= \int_{-\infty} ^{+\infty} \mathrm{d} (\Delta \tau) G^+_d(\Delta \tau) e^{-i\omega \Delta \tau},
\end{equation}
the transition rate can be written as 
\begin{equation}
\lambda^2\left((1-p)F_d(\omega)-pF_d(-\omega)\right).
\end{equation}
The $F_d(\omega)$ must include all the poles at the negative half of the vertical axis, so the pole at $i\varepsilon$ is not included. However, if we push the contour of the integration in the formula upwards by ${2\pi i}/{a}$, saying $\Delta \tau\rightarrow \Delta \tau+{2\pi i}/{a}$, the contribution from the pole at $i\varepsilon$ will be included. Thus we have
\begin{eqnarray}
\int_{-\infty} ^{+\infty} \mathrm{d} (\Delta \tau) G^+_d(\Delta \tau) e^{-i\omega \Delta \tau}&=&\int_{-\infty} ^{+\infty} \mathrm{d} (\Delta \tau) G^+_d(\Delta \tau+{2\pi i}/{a}) e^{-i\omega (\Delta \tau+{2\pi i}/{a})}\\ \nonumber
&-&\int_C \mathrm{d} (\Delta \tau) G^+_d(\Delta \tau) e^{-i\omega \Delta \tau},
\label{formula1}
\end{eqnarray}
where the integral $C$ is a circular contour with infinitesimal radius around the pole at $i\varepsilon$. From the formula of \eqref{Gd} we know that for $d\geq 3$ the Wightman function is periodic if $d$ is even, and anti-periodic if $d$ is odd, thus we have 
\begin{equation}
G^+_d(\Delta \tau+{2\pi i}/{a})=(-1)^d G^+_d(\Delta \tau),
\end{equation}
while for $d=2$ it will contribute a $-1$ in the $\log$ which can be left to the infinite constant. Using above two equations, we can directly obtain
\begin{equation}
\label{fd}
F_d(\omega)= \left[e^{2\pi \omega/a}-(-1)^d\right]^{-1}\int_C \mathrm{d} (\Delta \tau) G^+_d(\Delta \tau) e^{-i\omega \Delta \tau},
\end{equation}
where the $a/2\pi$ corresponds to the temperature predicted by the Unruh effect. The $\left[e^{2\pi \omega/a}-(-1)^d\right]^{-1}$ is the Bose-Einstein distribution for even $d$ and Fermi-Dirac distribution for odd $d$. The $\int_C \mathrm{d} (\Delta \tau)G^+_d(\Delta \tau) e^{-i\omega \Delta \tau}$ corresponds to the coefficients. For $d=2,3,4$, the coefficients are $1/\omega, 1/2, \omega/2\pi$ respectively.  Detailed results can be found in \cite{Sriramkumar:2002nt}. 

Here we give a brief discussion of eq.\eqref{fd}. It tells us that the Unruh effect predicts Bose-Einstein distribution in the even dimensions and Fermi-Dirac distribution in the odd dimensions, which seems to contradict our model since we are considering scalar fields. For simplicity, some textbooks explain the Unruh effect in the $(1+1)$-dimensional Rindler spacetime, since constant acceleration is a two-dimensional phenomenon and the addition of an extra dimension seems irrelevant. However, although the world line lies entirely in a two-dimensional plane, quantum field theory is not confined to that plane, but extends over the whole spacetime, so the extra dimension is not irrelevant. A more mathematical explanation was proposed by Ooguri \cite{Ooguri:1985nv}, which states the occurrence of Bose-Einstein or Fermi-Dirac spectrum depending on the dimension of the spacetime is because Huygens' principle is valid only in even dimensions. In fact, by global manifold embedding, the static observer in curved spacetime can be mapped into the accelerated observer in higher dimensional flat spacetime, and the temperature measured by both of them is the same \cite{Deser:1998xb}, however, we still cannot claim that the two are exactly equivalent, because quantum field theory is not equivalent in different dimensions.

Once we have $F_d(\omega)$, we can easily get $F_d(-\omega)$. The $F_d(\omega)$ is the transition rate of the qubit from an energy level $-\omega/2$ to $\omega/2$, while $F_d(-\omega)$ is the transition rate from $\omega/2$ to $-\omega/2$. If the value of $p$ is taken at the beginning just to reach thermal equilibrium with the Unruh temperature, i.e. when the Boltzmann distribution of the Unruh temperature is already satisfied, the transition rate should be zero. Therefore, if we directly take the value of $p$ as the Boltzmann factor at the Unruh temperature $\frac{1}{e^{2\pi \omega/a}+1}$ and let the transition rate be zero, then we have
\begin{equation}
F_d(-\omega)=e^{2\pi \omega/a}F_d(\omega),
\end{equation}
where the $e^{2\pi \omega/a}$ the Boltzmann factor expressing the relative population of the energy levels. The above equation holds for all dimensions.

With the results of $F_d(\pm \omega)$, we can directly calculate the transition rate in different dimensions. However, the decoherence rate is much more complicated because the integral is much harder to evaluate and we will find there are divergent terms that need to be renormalized. Fortunately, at this point we can already obtain our first theorem of this work:

\textbf{Theorem 1}: \emph{The real part of the decoherence rate is proportional to $\frac{1}{e^{2\pi \omega/a}-1}+1/2$ in even dimensions, and is proportional to $1/2$ in odd dimensions.}

\textbf{Proof}: The decoherence rate is proportional to
\begin{equation}
\int_{0} ^{+\infty} \mathrm{d} (\Delta \tau)G^+_d(\Delta \tau)e^{i\omega \Delta \tau}+\int_{-\infty} ^{0}\mathrm{d} (\Delta \tau)G^+_d(\Delta \tau)e^{-i\omega \Delta \tau}.
\end{equation}
It is hard to evaluate the result for all dimensions. However, due to the symmetry of the Wightman functions, we would have
\begin{equation}
\Re\left(\int_{0} ^{+\infty}\mathrm{d} (\Delta \tau)G^+_d(\Delta \tau)e^{i\omega \Delta \tau}\right)=\frac{1}{2}F_d(-\omega)
\end{equation}
and
\begin{equation}
\Re\left(\int_{-\infty} ^{0}\mathrm{d} (\Delta \tau) G^+_d(\Delta \tau)e^{-i\omega \Delta \tau}\right)=\frac{1}{2}F_d(\omega)
\end{equation}
For even dimensions
\begin{eqnarray}
\frac{1}{2}\left(F_d(-\omega)+F_d(\omega)\right)&=&\frac{1}{2}\left(\frac{e^{2\pi \omega/a}}{e^{2\pi \omega/a}-1}+\frac{1}{e^{2\pi \omega/a}-1}\right)\int_C \mathrm{d} (\Delta \tau) G^+_d(\Delta \tau) e^{-i\omega \Delta \tau}\\ \nonumber
&=&\left(\frac{1}{e^{2\pi \omega/a}-1}+\frac{1}{2}\right) \int_C \mathrm{d} (\Delta \tau) G^+_d(\Delta \tau) e^{-i\omega \Delta \tau} ,
\end{eqnarray}
while for odd dimensions
\begin{eqnarray}
\frac{1}{2}\left(F_d(-\omega)+F_d(\omega)\right)&=&\frac{1}{2}\left(\frac{e^{2\pi \omega/a}}{e^{2\pi \omega/a}+1}+\frac{1}{e^{2\pi \omega/a}+1}\right)\int_C \mathrm{d} (\Delta \tau) G^+_d(\Delta \tau) e^{-i\omega \Delta \tau}\\ \nonumber
&=&\frac{1}{2}\int_C \mathrm{d} (\Delta \tau) G^+_d(\Delta \tau) e^{-i\omega \Delta \tau} .
\end{eqnarray}
$\hfill\qedsymbol$

This theorem will directly lead to our corollary:

\textbf{Corollary}: \emph{The unitarity is always preserved in the time evolution of the Unruh-DeWitt detector in all dimensions.}

\textbf{Proof}: Considering the qubit in the initial state 
\begin{equation}
\rho_{0}=\begin{pmatrix} 
    p & \sqrt{p(1-p)} \\\nonumber

    \sqrt{p(1-p)} & 1-p
\end{pmatrix},
\end{equation}
after some time the density matrix will become
\begin{equation}
\begin{pmatrix} 
    p & \sqrt{p(1-p)} \\

    \sqrt{p(1-p)} & 1-p
\end{pmatrix}+
\begin{pmatrix} 
    \delta p & -\delta d \\

    -\delta d^* & -\delta p
\end{pmatrix},
\label{rhos2}
\end{equation}
where the $\delta p$ and $\delta d$ are proportional to the transition rate and decoherence rate respectively. Diagonalizing this density matrix we can have the von Neumann entropy
\begin{align}
S=-p_+\ln{p_+}-p_-\ln{p_-},
\end{align}
where 
\begin{eqnarray}
p_{\pm}&=&\frac{1\pm \sqrt{1+(8p-4)\delta p-8\sqrt{p(1-p)}\Re(\delta d)+4|\delta d|^2+4\delta p^2}}{2}
\\ \notag
       &= &\frac{1}{2}\pm \frac{1}{2}(1+(4p-2)\delta p-4\sqrt{p(1-p)}\Re(\delta d))+\mathcal{O}(\delta p^2, |\delta d|^2)
\end{eqnarray}
are the eigenvalues. Omitting the higher-order terms $\mathcal{O}(\delta p^2, \delta d^2)$, we get
\begin{eqnarray}
p_{+}&=&1-\left((1-2p)\delta p+2\sqrt{p(1-p)}\Re(\delta d)\right),
\\ \notag
p_{-}&=&(1-2p)\delta p+2\sqrt{p(1-p)}\Re(\delta d).
\label{popu}
\end{eqnarray}
To preserve the unitarity the $p_{-}$ must be non-negative. For the even dimensions we have
\begin{eqnarray}
p_{-}&\propto& (1-2p)\left((1-p)\bar{n}_b-p(\bar{n}_b+1)\right)+2p(1-p)\left(\bar{n}_b+\frac{1}{2}\right) \\ \nonumber
&=&(\bar{n}_b+1)p^2-2\bar{n}_bp+\bar{n}_b
\end{eqnarray}
where the $\bar{n}_b$ denotes the Bose-Einstein distribution $\frac{1}{e^{2\pi \omega/a}-1}$ that is always positive. We can easily verify that $p_{-}$ is non-negative because the discriminant $-4\bar{n}_b$ is always negative. 

Similarly for odd dimensions we have 
\begin{eqnarray}
p_{-}&\propto& (1-2p)\left((1-p)\bar{n}_f-p(1-\bar{n}_f)\right)+p(1-p) \\ \nonumber
&=&p^2-2\bar{n}_fp+\bar{n}_f,
\end{eqnarray}
where the $\bar{n}_f$ denotes the Fermi-Dirac distribution $\frac{1}{e^{2\pi \omega/a}+1}$ that satisfies $0<\bar{n}_f<1$. We can also have the discriminant $4\bar{n}_f(\bar{n}_f-1)$ is negative so the $p_{-}$ will always be non-negative.
 
$\hfill\qedsymbol$

\section{Results in low dimensions\label{section3}}

Although we have obtained the real part of the decoherence rate, we still do not have the imaginary part. We will find that for $d \geq 4$ there are divergence terms in the imaginary part, which must be renormalized by adding extra terms in the Hamiltonian. These divergence terms come from the contribution of the $i\varepsilon$. Note, however, that we cannot simply remove the $i\varepsilon$ contribution, as it may also contribute a real part factor. If the real part is removed, unitarity may be broken.

In principle we could solve each dimension separately, but this would be very complicated. Our strategy is to find recursive formulas between the different dimensions, so that the high-dimensional cases can be obtained directly from the low-dimensional results. In this section we first compute the cases of $d = 2$ and $d = 3$.

\subsection{$d=2$}

For the case of $d=2$, we have 
\begin{equation}
G^+_2(\Delta \tau)=-\frac{1}{2\pi}\log \left(\frac{2i}{a}\sinh \frac{a}{2}(\Delta \tau-i\varepsilon)\right).
\end{equation}
Here we have already renormalized the infinity constant comes from the dimensional regularization. If we compute $F(\pm \omega)$, the $\frac{2i}{a}$ in the $\log$ function can also be left to the infinity constant, since $\int_{-\infty}^{+\infty}e^{\pm i\omega \Delta \tau}\mathrm{d} \Delta \tau\times \text{const}$ is still proportional to $\delta({\omega})$ and does not contribute to the power spectrum for non-vanishing $\omega$. However, if the integration interval is $(-\infty, 0)$ or $(0,\infty)$, we need to keep it.

First we consider $\int_{0} ^{+\infty} \mathrm{d} (\Delta \tau)G^+_2(\Delta \tau)e^{i\omega \Delta \tau} $, we have
\begin{eqnarray}
\label{dta}
&-&\frac{1}{2\pi} \int_{0} ^{+\infty}\mathrm{d} (\Delta \tau) e^{i\omega \Delta \tau} \log \left(\frac{2i}{a}\sinh \frac{a}{2}(\Delta \tau-i\varepsilon)\right) \\ \nonumber
&=&-\frac{1}{2\pi} \int_{0} ^{+\infty}\mathrm{d} (\Delta \tau) e^{i\omega \Delta \tau} \log \left(\frac{2i}{a}\right) -\frac{1}{2\pi} \int_{0} ^{+\infty}\mathrm{d} (\Delta \tau) e^{i\omega \Delta \tau} \log \left(\sinh \frac{a}{2}(\Delta \tau-i\varepsilon)\right).
\end{eqnarray}
Using the equation
\begin{eqnarray}
\int_{0} ^{+\infty} \mathrm{d} (\Delta \tau) e^{i\omega \Delta \tau}= \pi\delta(\omega)+\frac{i}{\omega},
\end{eqnarray}
we can have the first term of the right hand side of \eqref{dta} is
\begin{eqnarray}
\label{log}
-\frac{1}{2\pi} \int_{0} ^{+\infty} \mathrm{d} (\Delta \tau) e^{i\omega \Delta \tau} \log \left(\frac{2i}{a}\right) =\frac{1}{4\omega}-\frac{i}{2\pi \omega}\log\left( \frac{2}{a} \right),
\end{eqnarray}
which contributes a real constant $\frac{1}{4\omega}$. In order to deal with the $\sinh(x)$ function, we need to use the following equation:
\begin{eqnarray}
\label{f1}
\sinh(x)=x\prod_{k=1}^{\infty}(k\pi-ix)(k\pi+ix)/(k\pi)^2,
\end{eqnarray}
so that the second term of the right hand side of \eqref{dta} can be written as
\begin{eqnarray}
-\frac{1}{\pi a}\bigg( \int_0^{+\infty}\mathrm{d}x e^{\frac{2\omega x}{a}i}\Big[\log(x-i\varepsilon)+\log\prod_{k=1}^{\infty}\big[k\pi-i(x-i\varepsilon)\big] \nonumber \\ 
 +\log\prod_{k=1}^{\infty}\big[k\pi+i(x-i\varepsilon)\big]-\log \prod_{k=1}^{\infty}k^2 \pi^2\Big]  \bigg),
\end{eqnarray}
where we have applied the variable transformation $\Delta \tau \rightarrow 2\omega x/a$. Next, we need to solve the four terms one by one.

For the first term, by virtue of the adiabatic-switching factor $s\rightarrow 0^+$, we may carry out the integration by part:
\begin{eqnarray}
&-&\frac{1}{\pi a} \int_0^{+\infty}\mathrm{d} x e^{i\frac{2\omega }{a}x-sx}\log(x-i\varepsilon)\nonumber \\
&=&\frac{i}{2\pi\omega}\left[\int_{0}^{\infty}-\frac{e^{i\frac{2\omega }{a}x-sx}}{x-i\varepsilon}\mathrm{d}x+e^{i\frac{2\omega }{a}x-sx}\log(x-i\varepsilon)\bigg|_{0}^{\infty}\right]\nonumber \\
&=&\frac{i}{2\pi\omega} \left[-E_1\left(-\varepsilon\left(\frac{2\omega}{a}+is\right)\right)e^{-\varepsilon\frac{2\omega}{a}}-\log(-i\varepsilon)\right]\nonumber \\
&=&\frac{i}{2\pi\omega}\left[-\left(-\gamma-\ln\left(\varepsilon\frac{2\omega}{a}\right)+ i\pi \right)-\log(-i\varepsilon)\right]\nonumber \\
&=&\frac{i}{2\pi\omega}\left[\gamma+\ln\left(\frac{2\omega}{a}\right)\right]+\frac{1}{4\omega},
\end{eqnarray}
where the $E_1(x)$ is the exponential integral which is a special function on the complex plane, and we have applied the formula
\begin{eqnarray}
\lim_{s\rightarrow 0_+}E_1(-x-is)=-Ei(x)+i\pi 
\end{eqnarray}
for $x>0$, and 
\begin{eqnarray}
Ei(x)=\gamma+\ln{|x|}-\sum_{k=1}^{\infty}\frac{(-1)^{k+1}(-x)^k}{kk!},
\end{eqnarray}
where $\gamma$ is the Euler–Mascheroni constant. Finally we obtain that there is a real constant $\frac{1}{4\omega}$. This constant and the $\frac{1}{4\omega}$ from eq.\eqref{log} will give the constant $1/2$ in the Theorem 1 for even dimension.

Now we move on to the second term. Here we need to use another equation:
\begin{eqnarray}
\label{f2}
\int_{\varepsilon}^{+\infty} \mathrm{d}y\frac{e^{-iy x}}{y(e^{\beta y}-1)}=-\ln{\left[\prod_{k=1}^{\infty}\varepsilon e^{\gamma}(\beta k+ix)\right]},
\end{eqnarray}
so that the second term can be written as
\begin{eqnarray}
\label{log2}
&-&\frac{1}{\pi a} \int_0^{+\infty}dx e^{\frac{2\omega x}{a}i}\log\prod_{k=1}^{\infty}\big[k\pi-i(x-i\varepsilon)\big]
\nonumber \\
&=&\frac{1}{\pi a}\left(\int_{0}^{+\infty}\mathrm{d}x e^{\frac{2\omega x}{a}i}\left[  \int_{\varepsilon}^{+\infty}\mathrm{d} y \frac{e^{iy x}}{y(e^{\beta y}-1)} +\log \prod_{k=1}^{\infty}\varepsilon e^{\gamma} \right]\right)
\nonumber \\
&=&\frac{1}{\pi a}\left(\int_{\varepsilon}^{+\infty}\mathrm{d} y \frac{\pi\delta(y+\frac{2\omega}{a})+\frac{i}{y+\frac{2\omega}{a}}}{y(e^{\pi y}-1)}+\log \prod_{k=1}^{\infty}\varepsilon e^{\gamma}\left( \pi\delta\left(\frac{2\omega}{a}\right)+\frac{i}{\frac{2\omega}{a}}\right)   \right)
\nonumber \\
&=&\frac{1}{\pi a}\left( \int_{\varepsilon}^{+\infty}\mathrm{d} y   \frac{i}{y\left(y+\frac{2\omega}{a}\right)(e^{\pi y}-1)}      +\frac{i}{\frac{2\omega}{a}}\log \prod_{k=1}^{\infty}\varepsilon e^{\gamma} \right).
\end{eqnarray}

Similarly the third term will be 
\begin{eqnarray}
&-&\frac{1}{\pi a} \int_0^{+\infty}\mathrm{d}x e^{\frac{2\omega x}{a}i}\log\prod_{k=1}^{\infty}\big[k\pi+i(x-i\varepsilon)\big]
\nonumber \\
&=&\frac{1}{\pi a}\left(\int_{0}^{+\infty}\mathrm{d}x e^{\frac{2\omega x}{a}i}\left[  \int_{\varepsilon}^{+\infty}\mathrm{d}y\frac{e^{-iy x}}{y(e^{\beta y}-1)} +\log \prod_{k=1}^{\infty}\varepsilon e^{\gamma} \right]\right)
\nonumber \\
&=&\frac{1}{\pi a}\left(\int_{\varepsilon}^{+\infty}\mathrm{d} y \frac{\pi\delta(y-\frac{2\omega}{a})-\frac{i}{y-\frac{2\omega}{a}}}{y(e^{\pi y}-1)}+\log \prod_{k=1}^{\infty}\varepsilon e^{\gamma}\left( \pi\delta\left(\frac{2\omega}{a}\right)+\frac{i}{\frac{2\omega}{a}}\right)   \right)
\nonumber \\
&=&\frac{1}{{2\omega}\left(e^{2\pi \omega/a}-1\right)} + \frac{1}{\pi a}\left(- \int_{\varepsilon}^{+\infty}\mathrm{d}y\frac{i}{y\left(y-\frac{2\omega}{a}\right)(e^{\pi y}-1)}      +\frac{i}{\frac{2\omega}{a}}\log \prod_{k=1}^{\infty}\varepsilon e^{\gamma} \right).
\end{eqnarray}
And the fourth term is 
\begin{eqnarray}
\frac{1}{\pi a}\int_0^{+\infty}\mathrm{d} x e^{\frac{2\omega x}{a}i} \log \prod_{k=1}^{\infty}k^2 \pi^2=\frac{i}{2\pi \omega}\log \prod_{k=1}^{\infty}k^2 \pi^2.
\end{eqnarray}

The $\int_{-\infty} ^{0}\mathrm{d}(\Delta \tau)G^+_2(\Delta \tau)e^{-i\omega \Delta \tau}$ can also be calculated in a similar way so we will not make a detailed derivation here. We can easily have 
\begin{equation}
-\frac{1}{2\pi} \int_{-\infty} ^{0}\mathrm{d}(\Delta \tau) e^{-i\omega \Delta \tau} \log \left(\frac{2i}{a}\right)=\frac{1}{4\omega}-\frac{i}{2\pi\omega}\log\left(\frac{2}{a}\right).
\end{equation}
For the other four terms, the first term we will have the real part is $-\frac{1}{4\omega}$, which is different from the case of $\int_{0} ^{\infty}\mathrm{d}(\Delta \tau)G^+_d(\Delta \tau)e^{i\omega \Delta \tau}$, while the sum of the remaining three terms is the same with the case of $\int_{0} ^{+\infty}\mathrm{d}(\Delta \tau) G^+_2(\Delta \tau)e^{i\omega \Delta \tau} d\Delta \tau$.

Adding all these terms together, we can obtain that the real part is proportional to $\frac{1}{e^{2\pi \omega/a}-1}+\frac{1}{2}$, and the imaginary part is
\begin{eqnarray}
\frac{i}{\pi \omega}(1+\ln \omega)&+&\frac{2i}{\pi a}\Big[ \int_{\varepsilon}^{\infty}\mathrm{d}y   \frac{1}{y\left(y+\frac{2\omega}{a}\right)(e^{\pi y}-1)} -\int_{\varepsilon}^{\infty}\mathrm{d}y   \frac{1}{y\left(y-\frac{2\omega}{a}\right)(e^{\pi y}-1)}\Big]
\nonumber \\
&+&\frac{2i}{\pi \omega} \log \prod_{k=1}^{\infty}  \varepsilon e^{\gamma} k \pi.
\end{eqnarray}
Using eq.\eqref{f2} again, we can write the last term as
\begin{eqnarray}
\frac{2i}{\pi \omega} \log \prod_{k=1}^{\infty}  \varepsilon e^{\gamma} k \pi=-\frac{2i}{\pi \omega}  \int_{\varepsilon}^{\infty} \frac{\mathrm{d}y}{y(e^{\pi y}-1)}.
\end{eqnarray}
All the three integral are divergent due to the $\varepsilon$, however, if we consider the behavior of $y\rightarrow 0$, we have
\begin{equation}
\frac{2i}{\pi a}\Big[\frac{1}{y\left(y+\frac{2\omega}{a}\right)(e^{\pi y}-1)}-\frac{1}{y\left(y-\frac{2\omega}{a}\right)(e^{\pi y}-1)}\Big] \longrightarrow \frac{2i}{\pi^2 \omega}\frac{1}{y^2}
\end{equation}
and we also have in the last term
\begin{equation}
-\frac{2i}{\pi \omega}\frac{1}{y(e^{\pi y}-1)}  \longrightarrow -\frac{2i}{\pi^2 \omega} \frac{1}{y^2},
\end{equation}
thus all the divergence would cancels out. 

\emph{In summary, in $d=2$ the real part of the decoherence rate is $\frac{1}{e^{2\pi \omega/a}-1}+\frac{1}{2}$, which is consistent with the Theorem 1, and the imaginary part is finite.}

\subsection{$d=3$}

Now we consider the case of $d=3$. The Wightman function
\begin{equation}
G^+_3(\Delta \tau)=\frac{a}{8\pi i}\frac{1}{\sinh \frac{a}{2}(\Delta \tau-i\varepsilon)},
\end{equation}
We still need to use the eq.\eqref{f1} so that the $\int_{0} ^{\infty}\mathrm{d}(\Delta \tau)G^+_3(\Delta \tau)e^{i\omega \Delta \tau}$ can be written as 
\begin{eqnarray}
\int_{0} ^{+\infty}\mathrm{d}(\Delta \tau)G^+_3(\Delta \tau)e^{i\omega \Delta \tau}&=&\frac{1}{4\pi i} \int_{0} ^{+\infty}\mathrm{d}x \frac{e^{\frac{2\omega x}{a}i}}{\sinh(x-i\varepsilon)}
\\ \nonumber
&=& \frac{1}{4\pi i} \int_{0} ^{+\infty}\mathrm{d}x \frac{e^{\frac{2\omega x}{a}i}}{(x-i\varepsilon)\prod_{k=1}^{\infty}(k\pi-i(x-i\varepsilon))(k\pi+i(x-i\varepsilon))/(k\pi)^2}
\end{eqnarray}
We can calculate the above equation directly and make use of the properties of exponential integral. Finally we have
\begin{eqnarray}
\int_{0} ^{+\infty}\mathrm{d}(\Delta \tau)G^+_3(\Delta \tau)e^{i\omega \Delta \tau}&=&\frac{i}{4\pi}\Bigg[ \sum_{k=1}^{\infty}\left( Ei\left(\frac{2\pi \omega}{a}k\right)e^{-\frac{2\pi \omega}{a}k}+Ei\left(-\frac{2\pi \omega}{a}k\right)e^{\frac{2\pi \omega}{a}k}   \right)
\nonumber \\
&+&\gamma+\ln\left(\frac{2\omega}{a}\varepsilon\right)  \Bigg]+\frac{1}{4}\frac{\frac{2\pi \omega}{a}}{e^{\frac{2\pi \omega}{a}}+1}.
\end{eqnarray}
The summation in the above equation is finite. Simialrly for $\int_{-\infty} ^{0}\mathrm{d}(\Delta \tau) G^+_d(\Delta \tau)e^{-i\omega \Delta \tau}$ we have
\begin{eqnarray}
\int_{-\infty} ^{0}\mathrm{d}(\Delta \tau)G^+_d(\Delta \tau)e^{-i\omega \Delta \tau} &=&-\frac{i}{4\pi}\Bigg[ \sum_{k=1}^{\infty}\left( Ei\left(\frac{2\pi \omega}{a}k\right)e^{-\frac{2\pi \omega}{a}k}+Ei\left(-\frac{2\pi \omega}{a}k\right)e^{\frac{2\pi \omega}{a}k}   \right)
\nonumber \\
&+&\gamma+\ln\left(\frac{2\omega}{a}\varepsilon\right)  \Bigg]+\frac{1}{4}\frac{1}{e^{\frac{2\pi \omega}{a}}+1}.
\end{eqnarray}
Thus
\begin{equation}
\int_{0} ^{\infty}\mathrm{d}(\Delta \tau)G^+_d(\Delta \tau)e^{i\omega \Delta \tau}+\int_{-\infty} ^{0}\mathrm{d}(\Delta \tau)G^+_d(\Delta \tau)e^{-i\omega \Delta \tau}=\frac{1}{4}.
\end{equation}
In $d=3$, the coefficient $\int_C \mathrm{d}(\Delta \tau) G^+_3(\Delta \tau) e^{-i\omega \Delta \tau}=\frac{1}{2}$, so the constant $\frac{1}{4}$ corresponds to the $\frac{1}{2}$ in odd dimensions in Theorem 1.

\emph{In summary, in $d=3$ the real part of the decoherence rate is proportional to $\frac{1}{2}$, as we proved in Theorem 1, and there is no imaginary term.}

\section{Recurrence formula for general $d$\label{section4}}

In this section we will describe how to get the high dimensional results from the low dimensional results. The formula of $G^+_2(\Delta \tau)$ is different from the case of $d\geq 3$, so we will consider it separately.

\subsection{From $d=2$ to $d=4$}
In this section we study the case of $d = 2$. In eq.\eqref{log2} we do the integration by part to convert $\log(x)$ to $1/x$. This gives us some clues: if we do the integration by part twice, we can get $1/x^2$, which might be related to $d=4$. We will show that given the $d$-dimensional result, we can directly get the $(d+2)$-dimensional case, and since we already have the cases of $d=2,3$, we can have the results for all $d$.

In the same way, we still consider the eq.\eqref{dta}, but this time we do the integration by part twice, and we have
\begin{eqnarray}
&-&\frac{1}{2\pi} \int_{0} ^{\infty}\mathrm{d}(\Delta \tau) e^{i\omega \Delta \tau} \log \left(\sinh \frac{a}{2}(\Delta \tau-i\varepsilon)\right)=-\frac{1}{\pi a} \int_0^{+\infty}\mathrm{d} x e^{i\frac{2\omega }{a}x-sx}\log\left(\sinh(x-i\varepsilon)\right)
\nonumber \\
&=&\frac{i}{2\pi \omega}\left[e^{i\frac{2\omega }{a}x-sx}\log\left(\sinh(x-i\varepsilon)\right) \Big|_0^{+\infty}-\int_0^{+\infty}\mathrm{d}x e^{i\frac{2\omega }{a}x-sx}\log\left(\coth(x-i\varepsilon)\right)\right]
\nonumber \\
&=& \frac{i}{2\pi \omega}\left[ -\log(-i \varepsilon) -\frac{1}{\frac{2\omega }{a}i}\left( \coth(x-i\varepsilon)\Big|_0^{+\infty}-\int_0^{+\infty}\mathrm{d}x e^{i\frac{2\omega }{a}x-sx} \frac{1}{\sinh(x-i\varepsilon)} \right)\right]
\nonumber \\
&=&\frac{i}{2\pi \omega}\left[ -\log(-i \varepsilon) -\frac{1}{\frac{2\omega }{a}i}\left( \frac{1}{i\varepsilon}-\int_0^{+\infty}\mathrm{d}x e^{i\frac{2\omega }{a}x-sx} \frac{1}{\sinh^2(x-i\varepsilon)} \right)\right]
\nonumber \\
&=& \frac{i}{2\pi \omega}\left[ -\log(\varepsilon)+\frac{a}{2\omega}\frac{1}{\varepsilon}\right]-\frac{1}{4\omega}+\frac{a}{4\pi \omega^2}\int_0^{+\infty}\mathrm{d} x e^{i\frac{2\omega }{a}x-sx} \frac{1}{\sinh^2(x-i\varepsilon)}.
\end{eqnarray}
We can find the last term of the above equation is exactly the case of $\int_{0} ^{+\infty}\mathrm{d}(\Delta \tau)G^+_4(\Delta \tau)e^{i\omega \Delta \tau}$ after we multiply by some coefficients. Previously we knew that the left side of the formula is finite, so there are two imaginary divergence terms in $\int_{0} ^{+\infty}\mathrm{d}(\Delta \tau)G^+_4(\Delta \tau)e^{i\omega \Delta \tau}$, which are $\log(\varepsilon)$ and $1/\varepsilon$. Similarly, we can also get
\begin{eqnarray}
&-&\frac{1}{2\pi} \int_{-\infty} ^{0}\mathrm{d}(\Delta \tau) e^{-i\omega \Delta \tau} \log \left(\sinh \frac{a}{2}(\Delta \tau-i\varepsilon)\right)
\nonumber \\
&=& \frac{i}{2\pi \omega}\left[ -\log(\varepsilon)-\frac{a}{2\omega}\frac{1}{\varepsilon}\right]-\frac{1}{4\omega}+\frac{a}{4\pi \omega^2}\int_{-\infty}^{0}dx e^{-i\frac{2\omega }{a}x+sx} \frac{1}{\sinh^2(x-i\varepsilon)}.
\end{eqnarray}
When we consider the decoherence rate in $d=4$, we will find the $1/\varepsilon$ is canceled out and only $\log(\varepsilon)$ remains, this is because the two divergence terms $\log(\varepsilon)$ and $1/\varepsilon$ are respectively the boundary terms generated from the twice integration by part, and the first one yields divergence terms $\log(\varepsilon)$ of the same sign, while the second ones $1/\varepsilon$ have the opposite, so that only the $\log(\varepsilon)$ exists in the case of $d = 4$.

\subsection{From $d$ to $d+2$}
For $d\geq3$, we have 

\begin{equation}
G^+_d(\Delta \tau)=\frac{\Gamma(d/2-1)}{4\pi^{d/2}}z^{2-d}=\frac{\Gamma(d/2-1)}{4\pi^{d/2}(2\pi i/a)^{d-2}}\frac{1}{\left(\sinh \frac{a}{2}(\Delta \tau-i\varepsilon)\right)^{d-2}},
\end{equation}
For the computational simplicity we omit the coefficients next, but note that the coefficients are real in even dimensions and imaginary in odd dimensions. Now we introduce the second theorem of this work.

\textbf{Theorem 2}: \emph{The imaginary part of the decoherence rate is finite in $d=2,3$ and contains a $\log(\varepsilon)$ divergence in $d=4$. For even dimensions $d\geq 6$, the divergence of the imaginary part can be expressed as
\begin{equation}
\log(\varepsilon)+\sum_{n=1}^{(d-4)/2}\frac{\alpha_n}{\varepsilon^{2n}},
\end{equation}
and for odd dimensions $d\geq 5$, the divergence can be expressed as
\begin{equation}
\sum_{n=1}^{(d-3)/2}\frac{\beta_n}{\varepsilon^{2n-1}},
\end{equation}
where the $\alpha_n$ and $\beta_n$ are the corresponding coefficients.
}

\textbf{Proof}: We have already calculated the cases of $d=2,3,4$. For $d\geq 3$, we first consider the case of 
\begin{equation}
\int_0^{+\infty}\mathrm{d}xe^{i\frac{2\omega}{a} x-sx} \frac{1}{\left(\sinh(x-i\varepsilon)\right)^{d-2}},
\end{equation}
where we have omitted the coefficients for simplicity. Do the integration by part twice the above formula can be written as
\begin{align}
&\frac{1}{i \frac{2 \omega}{a}}\left[\left.e^{i \frac{2 \omega}{a} x-s x} \frac{1}{\sinh (x-i \varepsilon)^{d-2}}\right|_{0} ^{+\infty}+(d-2) \int_{0}^{+\infty}\mathrm{d}x e^{i \frac{2 \omega}{a} x-s x} \frac{\cosh (x-i \varepsilon)}{\sinh (x-i \varepsilon)^{d-1}}\right]
\nonumber \\
&= \frac{1}{(-i)^{d-1} \frac{2 \omega}{a}} \frac{1}{\varepsilon^{d-2}}+\frac{(d-2)}{i \frac{2 w}{a}}\Bigg[\frac{1}{i \frac{2 w}{a}}\bigg(\left.e^{i \frac{2 \omega}{a} x-s x} \frac{\cosh (x-i \varepsilon)}{\sinh (x-i\varepsilon )^{d-1}}\right|_{0} ^{+\infty}
\nonumber \\
&+\int_{0}^{+\infty}\mathrm{d}x e^{i \frac{2 \omega}{a} x} \frac{(d-2) \sinh \left(x-i\varepsilon \right)^{2}+(d-1)}{\sinh (x-i \varepsilon)^d}\bigg)\Bigg] 
\nonumber \\
& = \frac{1}{(-i)^{d-1} \frac{2 w}{a}} \frac{1}{\varepsilon^{d-2}}+  \frac{(d-2)}{\left(\frac{2 \omega}{a}\right)^{2}(-i)^{d-1}} \frac{1}{\varepsilon^{d-1}}-\frac{(d-2)^{2}}{\left(\frac{2 \omega}{a}\right)^{2}} \int_{0}^{+\infty} e^{i \frac{2 \omega }{a} x} \frac{1}{\sinh (x-i \varepsilon)^{d-2}}\mathrm{d} x 
\nonumber \\
& -\frac{(d-2)(d-1)}{\left(\frac{2 w}{a}\right)^{2}} \int_{0}^{+\infty} e^{i \frac{2\omega}{a} x} \frac{1}{\sinh (x-i \varepsilon)^{d}} \mathrm{d} x.
\end{align}
Thus we have
\begin{align}
 \int_{0}^{+\infty} e^{i \frac{2 \omega}{a} x} \frac{(d-2)(d-1)}{\sinh (x-i \varepsilon)^{d}} d x & =-\left[\left(\frac{2 \omega}{a}\right)^{2}+(d-2)^{2}\right] \int_{0}^{+\infty} e^{i \frac{2 \omega}{a} x} \frac{1}{\sinh (x-i \varepsilon)^{d-2}} d x 
\nonumber \\
& +\frac{2 w}{a} \frac{1}{(-i)^{d-1}} \frac{1}{\varepsilon^{d-2}}+\frac{(d-2)}{(-i)^{d-1}} \frac{1}{\varepsilon^{d-1}},
\end{align}
where the left hand side corresponds to the $(d+2)$-dimensional case, and the first term of the right hand side corresponds to the $d$-dimensional case.

Similarly we also have
\begin{align}
 \int_{-\infty}^{0} e^{-i \frac{2 w}{a} x} \frac{(d-2)(d-1)}{\sinh (x-i \varepsilon)^{d}} d x & =-\left[\left(\frac{2 w}{a}\right)^{2}+(d-2)^{2}\right] \int_{-\infty}^{0} e^{-i \frac{2 w}{a} x} \frac{1}{\sinh (x-i \varepsilon)^{d-2}} d x 
\nonumber \\
& +\frac{2 w}{a} \frac{1}{(-i)^{d-1}} \frac{1}{\varepsilon^{d-2}}-\frac{(d-2)}{(-i)^{d-1}} \frac{1}{\varepsilon^{d-1}}.
\end{align}
When we consider the decoherence rate, we can find that for $d\geq 3$ the $(d+2)$-dimensional case corresponds to the $d$-dimensional result plus an $1/\varepsilon^{d-2}$ divergence term, while the $1/\varepsilon^{d-1}$ cancels out. The sum of the above two equations is the recurrence formula of decoherence rate. We have given the cases of $d=2,3,4$, so the theorem is proved.
$\hfill\qedsymbol$

Since we have obtained the transition rate and the decoherence rate, we can add the renormalization term in $\hat{H}_0$ to eliminate the corresponding divergence. Related discussions can also be found in \cite{Kaplanek:2019dqu,Ali:2020gij}.  Note, however, that we cannot naively eliminate the $i\varepsilon$. Although this will cure the divergence, it will also subtract the real term, which may break the unitarity.

\section{Conclusions and outlook\label{section5}}

In this work, we study the interaction of a qubit with an initial pure state as an Unruh-DeWitt detector coupled to a free massless scalar field in arbitrary dimensions. Using the perturbation method, we calculate the transition rate and the decoherence rate of the qubit. Our main results are two theorems and one corollary that we proved in the paper. We list them below:

\begin{itemize}
  \item \emph{Real part of the decoherence rate.} The real part of the decoherence rate is proportional to $\frac{1}{e^{2\pi \omega/a}-1}+1/2$ in even dimensions, and is proportional to $1/2$ in odd dimensions.

\item \emph{Imaginary part of the decoherence rate.} The imaginary part of the decoherence rate is finite in $d=2,3$ and contains a $\log(\varepsilon)$ divergence in $d=4$. For even dimensions $d\geq 6$, the divergence of the imaginary part can be expressed as
\begin{equation}
\log(\varepsilon)+\sum_{n=1}^{(d-4)/2}\frac{\alpha_n}{\varepsilon^{2n}},\nonumber
\end{equation}
and for odd dimensions $d\geq 5$, the divergence can be expressed as
\begin{equation}
\sum_{n=1}^{(d-3)/2}\frac{\beta_n}{\varepsilon^{2n-1}}.\nonumber
\end{equation}

  \item \emph{Unitarity of the Unruh-DeWitt detector.} The unitarity is always preserved in the time evolution of the Unruh-DeWitt detector in all dimensions.
\end{itemize}

We conclude with a brief discussion of possible applications of this work. The transition rate and the decoherence rate are both important physical observable quantities. They directly relate the nature of the detector and the quantum field theory, which may provide clues for future experiments, especially in the future spacebased experiments \cite{Anastopoulos:2021llw}. Recently it was found that the decoherence rate can be used to measure the Unruh effect \cite{Nesterov:2020exl}. We can also study the exchange of energy and information during the interaction between the detector and quantum field theory \cite{Xu:2021buk,Xu:2022juv,Xu:2021ihm}, which may help us to understand the feedback and back-reaction obtained from the measurements, as well as the connection between gravity and quantum theory. In \cite{Xu:2022juv} we also investigated the decoherence of the qubit coupled to a thermal quantum field theory in a cavity, and the results are also consistent with our Theorem 1. Of course, when we consider the thermal state instead of the Unruh effect, the Fermi-Dirac distribution does not appear in odd dimensions. Interested readers can find a more detailed discussion in \cite{Takagi:1986kn}. 

Here we make a brief comparison between our work and \cite{Audretsch:1995qt}, where the authors also studied the decoherence for the Unruh-DeWitt detector, although not in arbitrary dimensions. In addition to the dimension, there is another major difference: the authors of \cite{Audretsch:1995qt} considered the decoherence with continuous measurement, while in our work we do not. The case with continuous measurement considers finite time integration, while the case without continuous measurement considers infinite time integration. When the time tends to infinity, the former reduces to the latter. In the present work we study the decoherence rate and thermalization rate, where the integration range is always infinite, while in \cite{Xu:2022juv} we investigated a situation similar to the \cite{Audretsch:1995qt}. Interested readers can find a more detailed discussion in these papers. Also, the decoherence in \cite{Audretsch:1995qt} is mainly from the transition between the ground state and the excited state, which is actually more like the thermalization in this paper, because the thermalization process from the pure state to the mixed state can also be decoherent. The initial state in our work is the linear superposition of the ground and excited states, and we can calculate not only the evolution of diagonal elements in the density matrix, but also the off-diagonal elements, so that we can understand the Unruh effect on the qubit more clearly.

On the other hand, many quantum gravity models exhibit dynamical dimensional reduction in the microscopic region \cite{Carlip:2017eud}, and the corrections to the two-point function may leave characteristic fingerprints that can be captured by detectors.   Since we have obtained the results for all dimensions, the decoherence rate can also be used as an effective quantity to probe spacetime on small scales. There are some works that has made similar attempts \cite{Alkofer:2016utc,Saueressig:2021pzy}. Because our results show that the response of the Unruh-DeWitt detector varies with dimensions, there are more interesting issues to investigate here, particularly quantum field theory and quantum gravity in non-integer dimensions \cite{Akkermans:2010dz,Eckstein:2020gjd}, and the corresponding response functions of the Unruh-DeWitt detector. Furthermore, we can also study the decoherence of qubit placed in the curved spacetime with black holes, even the decoherence of black holes themselves \cite{Demers:1995tr}. We will pursue these directions in our future work.

\section*{Acknowledgements}
Hao Xu thanks Yuan Sun for useful discussions. He also thanks National Natural Science Foundation of China (No.12205250) for funding support. 

%\appendix

\bibliographystyle{JHEP}

\bibliography{draft-bibliography}

\end{document}